\documentclass[lettersize,journal]{IEEEtran}
\usepackage{amsmath,amsfonts}
\usepackage{algorithmic}
\usepackage{algorithm}
\usepackage{array}
\usepackage[caption=false,font=normalsize,labelfont=sf,textfont=sf]{subfig}
\usepackage{textcomp}
\usepackage{stfloats}
\usepackage{url}
\usepackage{verbatim}
\usepackage{graphicx}
\usepackage{pifont}
\usepackage{multirow}
\usepackage{booktabs}
\usepackage{arydshln} 

\usepackage{xcolor}
\usepackage{amssymb}
\definecolor{ablgray}{gray}{0.68}

\newcommand{\graytext}[1]{{\color{ablgray}#1}}

\usepackage{cite}
\usepackage{siunitx}
\usepackage[colorlinks = true,
            linkcolor = blue,
            urlcolor  = blue,
            citecolor =  green,
            anchorcolor = blue]{hyperref}

\begin{document}

\newcommand{\vModelName}{\textit{AudioLDM 2}}

\title{FlowSep 2: Self-Supervised Flow Matching for Language-Queried Audio Source Separation}

\author{Yi Yuan, Xubo Liu, Haohe Liu, Xiyuan Kang, Mark D. Plumbley,  Wenwu Wang  \\ 
\thanks{Yi Yuan,Xubo Liu, Haohe Liu Xiyuan Kang, and Wenwu Wang are with the School of Computer Science and Electronic Engineering, University of Surrey, Guildford, UK. Email: \{yi.yuan, xubo.liu, haohe.liu, xk00063, w.wang\}@surrey.ac.uk.}
\thanks{ Mark D. Plumbley is with the Department of Informatics, King's College London, London, UK. Email: mark.plumbley@kcl.ac.uk.}

\thanks{Demos are available at~\url{https://audio-agi.github.io/Flowsep2-demo/}.}
}

\maketitle

 \begin{abstract}
Language-queried audio source separation (LASS) aims to extract target sources from audio mixtures according to natural language descriptions, offering a flexible and scalable interface for audio source separation. However, most existing LASS methods rely on discriminative, mask-based models, which estimate masks from the input mixture. These methods often over-suppress target sounds or fail to fully separate them, especially when multiple sound events strongly overlap in complex acoustic scenes. In this work, we propose FlowSep2, a text-conditioned flow-matching generative model for LASS. Instead of directly predicting a separation mask, FlowSep2 learns to generate the target source representation from Gaussian noise in a latent space, conditioned on both the mixture representation and the text query. Specifically, we employ rectified flow matching with a Diffusion Transformer backbone. 
We further incorporate Self-Flow, a self-supervised flow-matching paradigm, into our LASS framework. By encouraging semantically structured latent representations under the generative objective, Self-Flow improves the model’s ability to separate target sources according to text queries. Experiments on multiple LASS benchmarks show that FlowSep2 achieves state-of-the-art performance and demonstrates enhanced sound separation results in challenging scenarios with overlapping sound events. 
\end{abstract}

\begin{IEEEkeywords}
sound separation, language-queried audio
source separation (LASS), natural language processing, diffusion transformers, rectified flow matching
\end{IEEEkeywords} 

\section{Introduction}

\IEEEPARstart{S}ound separation is a fundamental task in computational auditory scene analysis~(CASA)~\cite{liu2019divide}, which aims to decompose audio mixtures into individual source signals. Benefiting from large-scale multi-modal data, recent systems have demonstrated strong generalization ability across various separation scenarios, including universal audio separation~\cite{universal1,universal2,liu2024audio}, speech separation~\cite{luo2019convtasnet,speech2}, and music source separation~\cite{music1,music2}. 
These systems have significantly broadened the applicability of sound separation technologies, supporting tasks such as automatic audio editing~\cite{audioedit,audioedit2}, multimedia retrieval~\cite{clip}, assistive listening~\cite{augmentlisten}, and speech enhancement in noisy environments~\cite{kong2021speech}. 

In recent years, increasing attention has been devoted to language-queried audio source separation~(LASS)~\cite{liu2022separate}. By leveraging textual queries as conditioning signals, LASS enables users to extract desired audio events through intuitive and flexible instructions. Compared with label-queried methods~\cite{veluri2023realtime,delcroix2022soundbeam,wang2022timestamp}, or audio-visual-queried methods~\cite{audio-visual-1,audio-visual-2}, LASS removes predefined category constraints and does not rely on additional modalities. This substantially broadens the range of separable sound sources and enables more convenient and flexible interaction with source separation systems.

 Many previous models for LASS, such as LASS-Net~\cite{liu2022separate} and AudioSep~\cite{audiosep}, primarily use discriminative methods~\cite{veluri2023realtime}, which predict time-frequency masks from spectrograms. Despite giving promising results for universal audio sources, these systems may face challenges when dealing with overlapping sound events~\cite{universal1}. In particular, the masks generated by these models may be excessive or insufficiently selective~\cite{luo2019convtasnet}, leading to artifacts such as spectral holes or incomplete separation~\cite{speech2}. These limitations restrict their effectiveness in real-world applications that involve diverse and dynamic acoustic environments.

Recent advances in diffusion-based generative models~\cite{ddim,DDPM,valle2025fugatto} and large-scale audio-language datasets~\cite{audioset,wavcaps,autoacd,sound_vecaps} have facilitated the development of generative approaches~\cite{hai2024dpmtse} for LASS. Unlike discriminative approaches that directly estimate spectrogram masks, generative models~\cite{audioldm,audioldm2} learn the underlying data distribution and generate outputs conditioned on textual guidance. This avoids the masking errors caused by heavily overlapping sound events.
Among recent generative formulations, Rectified Flow Matching (RFM)~\cite{liu2023rectifiedflow} has attracted increasing attention due to its theoretical simplicity and efficient sampling properties. RFM learns a velocity field that transports samples from the noise distribution to the data distribution along linear flow trajectories. This formulation reduces sampling complexity while preserving strong generative capacity, making it particularly suitable for separation systems.

Our previous work, FlowSep~\cite{flowsep}, is one of the early works that introduced a generative network for LASS by leveraging RFM in audio feature latent space. FlowSep learns to generate the latent representation of the target source conditioned on the mixture audio and the textual query. This model improves overall separation performance, particularly for mixtures containing heavily overlapping sound events. 


In this work, we advance FlowSep toward a more scalable generative framework for LASS, leading to enhanced separation performance and stronger semantic consistency between the generated audio and textual queries. We first enhance the model architecture and training scale to improve open-domain generalization capability. In addition, inspired by Self-Flow~\cite{self-flow}, we adapt self-supervised flow representation learning to conditional source separation. Specifically, we introduce a representation alignment objective that encourages latent representations to preserve consistent semantic information. We refer to this objective as semantic representation alignment. The proposed system, named as \textbf{FlowSep2}, significantly improves separation quality and text-audio semantic alignment across diverse benchmarks. The main contributions of this work are summarized as follows:

\begin{itemize}

\item We establish a generative paradigm for LASS through RFM, enabling target sources to be directly generated in the latent space conditioned on mixture audio and natural-language queries.

\item We extend the generative separation framework with self-supervised representation alignment, improving the semantic correspondence between learned representations and language-specified target sources.

\item We develop a scalable DiT-based architecture with an end-to-end waveform encoder-decoder and investigate model scaling to enhance global modeling capability for complex acoustic scenes.

\item We scale the training of FlowSep2 to over $5{,}000$ hours of audio-language data, achieving strong generalization across diverse sound separation benchmarks and acoustic domains.

\item Through comprehensive evaluations and extensive ablation studies, we analyze the effects of generative formulation, representation alignment learning, backbone architecture, and model scaling, providing insights into the key design choices for generative LASS systems.

\end{itemize}

The remainder of the paper is organised as follows. Section~\ref{sec:related-work} discusses related work. Section~\ref{sec: FlowSep2} presents the details of the proposed FlowSep2 model. Section~\ref{sec: dataset} discusses the datasets and metrics used for performance benchmarking. Section~\ref{sec: experiment} discusses the model setup including training and inference. Section~\ref{sec: results} presents the experimental results and performance analysis. Section~\ref{sec: ablation} provides ablation studies to demonstrate the contributions of model components. Finally, Section~\ref{sec: conclusion} concludes the paper with a discussion of future work.

\section{Related Work}
\label{sec:related-work}

\subsection{Audio Source Separation}

The objective of an audio source separation system is to decompose an audio mixture into its constituent source signals without requiring prior knowledge of the underlying mixing process. Deep learning approaches achieve state-of-the-art (SoTA) performance, and can be broadly divided into time-domain and frequency-domain methods. Time-domain models operate directly on raw waveforms, with representative architectures including WaveUNet~\cite{Wave-U-Net}, ConvTasNet~\cite{luo2019convtasnet}, and Demucs~\cite{defossez2021hybrid}. In contrast, frequency-domain methods~\cite{speech2,hershey2016deepclustering,wang2018multichannel} perform separation in the time-frequency domain by estimating unmixing filters or masks and applying them to mixture spectrograms to recover individual sources. More recently, hybrid training strategies incorporating joint time and frequency domain supervision have been proposed to improve separation robustness and reconstruction quality~\cite{wan2023multiloss}. Beyond direct signal mapping, representation learning approaches~\cite{wang2018multichannel,hershey2016deepclustering} learn discriminative embedding spaces in which sources are separated through clustering objectives, demonstrating strong performance, particularly in speech separation.

More recently, research has shifted toward Universal Sound Separation (USS)~\cite{yu2017pit,wisdom2020mixit}, which aims to separate arbitrary sound sources encountered in open-world acoustic environments. Compared with conventional source separation, USS is considerably more challenging due to the diversity and unpredictable nature of real-world sound events. Permutation Invariant Training (PIT)~\cite{yu2017pit} enables supervised learning for mixtures containing arbitrary sources by matching predictions with reference signals during training. Building upon this idea, Mixture Invariant Training (MixIT)~\cite{wisdom2020mixit} further removes the need for isolated ground-truth sources by exploiting noisy real-world mixtures in an unsupervised learning framework, achieving competitive performance under realistic acoustic conditions. Nevertheless, both PIT and MixIT approaches typically rely on post-processing or source selection modules to associate separated signals with semantic categories. This class-dependent design limits their scalability in open-domain scenarios, motivating the development of more flexible separation frameworks that incorporate semantic guidance directly into the separation process.

\subsection{Query-based Sound Separation}

Query-based sound separation (QSS) aims to separate a specific source from an audio mixture, based on auxiliary query information. Depending on the modality and form of the query signal, existing approaches can be broadly categorized into audio-queried, label-queried, and language-queried paradigms. 

\subsubsection{Audio-Queried Sound Separation}
Audio-queried sound separation conditions the separation model on reference audio recordings of the target source. One-shot and few-shot approaches use one or a few examples to extract acoustically similar sources from mixtures by encoding the references into query embeddings~\cite{gfeller2021oneshot, wang2022fewshot}. To reduce the reliance on labeled single-source data, weakly supervised methods construct training mixtures using anchor segments obtained from sound event detection models~\cite{kong2023universal, chen2022zeroshot}. Although these methods generalize well to unseen sound sources, they require reference audio during inference, limiting their practicality in real-world applications.

\subsubsection{Label-Queried Sound Separation}
Label-queried sound separation conditions the separation model on discrete sound class labels~\cite{veluri2023realtime, delcroix2022soundbeam,wang2022timestamp}. Unlike audio-queried methods, it eliminates the need for reference recordings, making interaction more convenient. However, predefined label sets limit scalability to open-domain scenarios and cannot capture richer semantic information such as temporal relationships, spatial context, or multiple concurrent events. Expanding the supported vocabulary typically requires retraining or continual learning, increasing computational cost and maintenance complexity.


\subsubsection{LASS}
Recently, LASS has extended query-based sound separation by conditioning separation models on natural language descriptions, enabling more expressive semantic guidance. Early work, such as LASS-Net~\cite{liu2022separate}, introduced end-to-end language-conditioned separation using paired audio--text data. To address the limited availability of annotated audio--text datasets, AudioSep~\cite{audiosep} leverages pretrained multimodal models such as Contrastive Language-Audio Pretraining (CLAP)~\cite{clap} to align audio and text representations, enabling zero-shot separation from either text or audio queries. In addition, OmniSep~\cite{omnisep} unifies text, image, and audio-queried sound separation within a single framework, allowing target sources to be specified using multiple query modalities. PromptSep~\cite{promptsep} employs a conditional diffusion model for generative audio separation and supports both sound extraction and removal using text and vocal prompts. More recently, SAM-Audio~\cite{sam-audio} introduces a general audio separation foundation model based on a flow-matching transformer, supporting text, visual, and temporal-span prompts for separating speech, music, and general sounds.

Despite these advances, many existing LASS systems still rely on masking-based discriminative architectures~\cite{kilgour2022textdriven,tzinis2023oct}, which can struggle with heavily overlapping sources and acoustically complex sound events. In this work, we focus on the language-queried setting and generative models, in which natural language descriptions provide flexible and expressive semantic conditions for specifying target sound events.

\subsection{Text-to-Audio Generation}
Text-to-audio generation has emerged as an important area in audio synthesis, enabling users to guide generative models with specific constraints or conditions using text prompts to produce desired outcomes~\cite{controlnet}. As one of the earlier models, AudioLDM~\cite{audioldm} employs the CLAP model to generate the embeddings of audio and text, and learns a latent diffusion model (LDM) for audio synthesis from latent vectors produced by a variational autoencoder (VAE)~\cite{kingma2014auto}. 

Beyond stochastic diffusion processes, recent studies have investigated flow-based generative formulations, such as RFM~\cite{liu2023rectifiedflow}, which learn deterministic trajectories that directly connect noise and target data distributions. 
Compared with conventional diffusion models, RFM-based approaches enable more efficient sampling with fewer inference steps while retaining strong generative capabilities. Recent audio generation models have further combined flow-based objectives with Transformer-based architectures. For example, Tango-Flux~\cite{tangoflux} integrates flow matching with a Diffusion Transformer (DiT) backbone, achieving state-of-the-art performance on text-to-audio benchmarks.

\subsection{Semantic Representation Alignment in Generative Models} 
Recent studies have increasingly highlighted the role of semantic representation learning in generative modeling. Representation Autoencoders (RAE)~\cite{rae} replace conventional VAE encoders with pretrained representation models, leading to improved training stability and generative fidelity. Similarly, Representation Alignment (REPA)~\cite{repA} aligns intermediate representations of generative models with those from pretrained semantic encoders, allowing the models to leverage richer semantic information from external representations.

More recently, Self-Flow~\cite{self-flow} introduces a self-supervised approach that enhances semantic representations within the generative training process itself. Rather than relying on external representation models, Self-Flow employs an Exponential Moving Average (EMA) teacher-student framework, in which two noisy views with distinct flow trajectories are generated from the same training samples. Enforcing consistency between the representations extracted from these views enables the model to learn semantically meaningful latent features while jointly optimizing the generative objective. Inspired by Self-Flow, we adapt this representation learning paradigm to LASS; the details of the proposed objective are presented in Section~\ref{sec: diffusion}.

\begin{figure*}[htbp]
    \centering
    \includegraphics[width=1.0\linewidth]{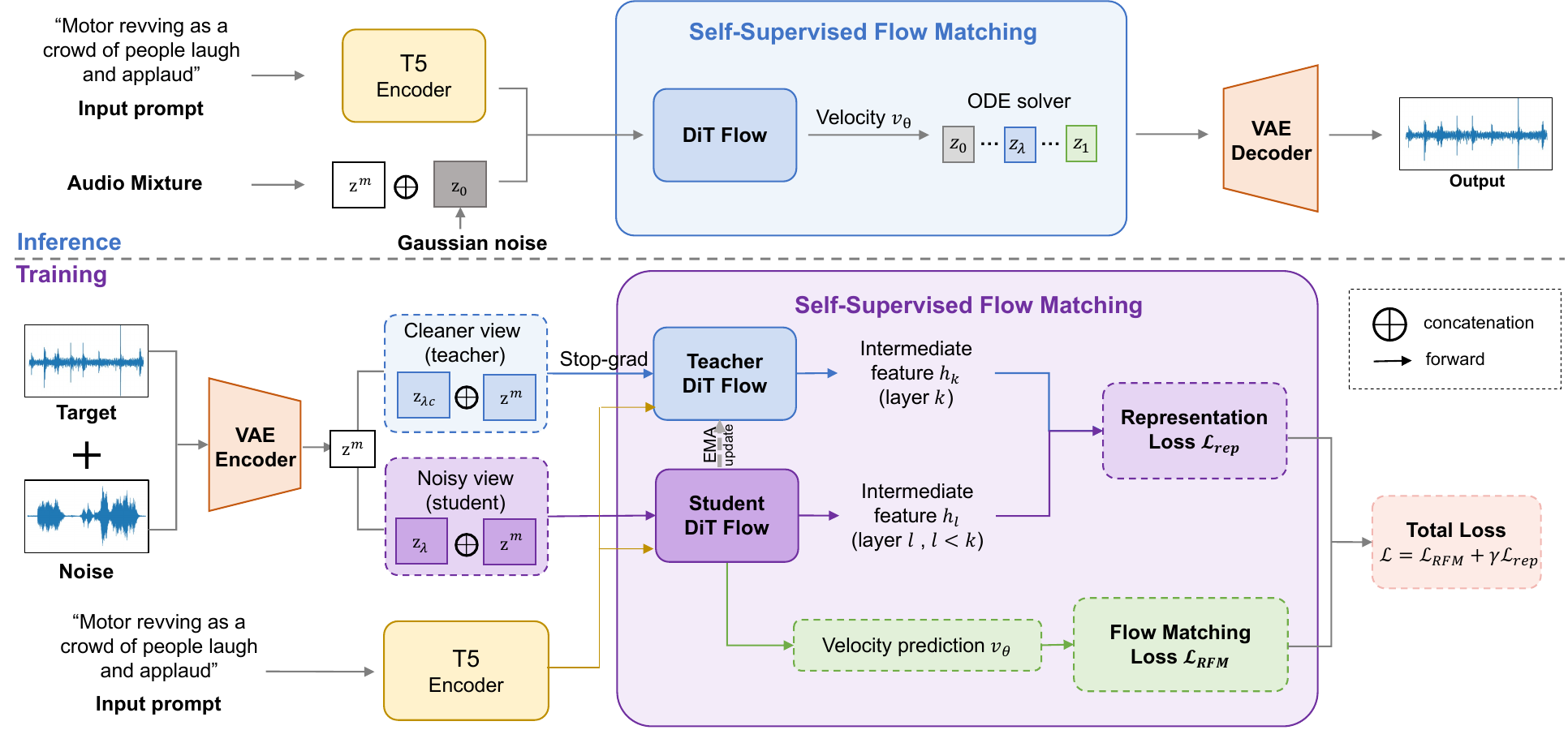}
    \caption{The overall architecture of FlowSep2. Here, $\mathbf{z}^{m}$ denotes the mixture latent, $\mathbf{z}_0$ denotes the Gaussian noise, $\mathbf{z}_1$ denotes the target latent, and $\mathbf{z}_{\lambda}$ represents the intermediate latent along the flow trajectory. During training, the mixture latent $\mathbf{z}^{m}$ is concatenated with the intermediate latent before being fed into the DiT flow module and the teacher receives a cleaner intermediate latent $\mathbf{z}_{\lambda c}$ to provide representation-level supervision. The teacher network is updated using EMA and the student network is optimized using both the flow matching loss $\mathcal{L}_{\mathrm{RFM}}$ and the representation loss $\mathcal{L}_{\mathrm{rep}}$.}
    \label{fig:overview}
\end{figure*}

\subsection{Sound Separation with Generative Models}


Generative modeling has emerged as a promising alternative to conventional masking-based separation frameworks~\cite{hai2024dpmtse}. Rather than directly estimating the source masks from the input mixture, generative approaches perform separation by synthesizing the waveform of the target source. FlowSep~\cite{flowsep} represents our initial effort to explore generative modeling for audio source separation. Experimental results demonstrate that generative modeling provides a viable and effective alternative to discriminative masking-based approaches for language-guided audio source separation.

Recently, large-scale industrial systems have also investigated generative approaches for universal sound separation. SAM-Audio~\cite{sam-audio}, for example, introduces a foundation-level framework for audio segmentation and separation trained on large-scale proprietary datasets. Although such systems achieve strong performance in open-domain settings, their dependence on massive private datasets and undisclosed training pipelines limits reproducibility and prevents systematic analysis by the research community. More broadly, existing generative separation methods still face challenges in robust open-domain generalization, particularly in complex acoustic scenes with heavily overlapping sound events. These challenges motivate our work on improving semantic modeling, architectural scalability, and large-scale training for language-guided sound separation.

\section{Proposed Method} 
\label{sec: FlowSep2}


The proposed FlowSep2 model is illustrated in Figure~\ref{fig:overview}, which is composed of several modules, including the text encoder~(e.g. FLAN-T5) to obtain the embeddings of the input text queries, the audio encoder~(e.g. Stable-Audio~\cite{stableaudio} VAE encoder) to obtain the embeddings of the mixture audios, the feature generator built on Self-Flow, and the VAE decoder for reconstructing the target audio, as detailed below.

\subsection{Text and Audio Encoders}
\label{sec: condition}
\subsubsection{Text Encoder} For the target text prompts, we use the pre-trained FLAN-T5~\cite{t5} as the text encoder to extract the text feature. Compared to contrastive language pretraining models, such as CLIP~\cite{clip} and CLAP~\cite{clap}, the FLAN-T5 encoder captures both semantic meaning~\cite{tango} and temporal structures~\cite{tclap} from textual prompts, showing enhanced performance in extracting semantic information for generation tasks~\cite{tango,reaudioldm}. In detail, for each query $\boldsymbol{c}$, the text embedding $\boldsymbol{E}$ is obtained as: 
\begin{equation}
\label{eqa:1}
    \boldsymbol{E} = \textit{f}_{\text{T5}}(\boldsymbol{c})
\end{equation}
where $\textit{f}_{\text{T5}}(\cdot)$ is the FLAN-T5 text encoder~\cite{t5}. 

\subsubsection{Audio Embedding}

For audio embeddings, we employ a pre-trained VAE~\cite{kingma2014auto} from Stable-Audio~\cite{stableaudio} to map waveform signals into a compact latent space. Given an input audio signal $a$, the latent representation $\boldsymbol{z}_1$, which also serves as the target representation, is obtained through the VAE encoder. During inference, the predicted denoised latent representation $\hat{\boldsymbol{z}}_1$ is mapped back to waveform space through a VAE decoder: 

\begin{equation}
\boldsymbol{z}_1 = f_{\text{VAE}}(a), \qquad
\hat{a} = D_{\text{VAE}}(\hat{\boldsymbol{z}}_1)
\end{equation}
\noindent where $f_{\text{VAE}}$ denotes the VAE encoder and $D_{\text{VAE}}$ denotes the VAE decoder.

\subsection{Audio Feature Generation}
\label{sec: diffusion}

Traditional diffusion-based generative models, such as Denoising Diffusion Probabilistic Models~(DDPMs)~\cite{DDPM}, generate samples through iterative denoising from a Gaussian prior. These methods gradually transform noise into data using hundreds of discrete denoising steps. RFM~\cite{liu2023rectifiedflow} provides an alternative formulation by learning a transformation between the noise and data distributions. Instead of predicting noise at discrete time steps, RFM directly learns a velocity field that transports samples from the noise distribution to the target data distribution. 

\subsubsection{Rectified Flow Matching}
RFM models the linear transformation path between a Gaussian noise latent $\boldsymbol{z}_0 \sim \mathcal{N}(\mathbf{0}, \mathbf{I})$ and the target latent representation $\boldsymbol{z}_1$. A continuous flow variable $\lambda \in [0,1]$ is introduced to interpolate between these two states. The intermediate noisy representation $\boldsymbol{z}_{\lambda}$ is defined as
\begin{equation}
\boldsymbol{z}_{\lambda} =
\left(1 - (1-\sigma)\lambda\right)\boldsymbol{z}_{0}
+ \lambda\boldsymbol{z}_{1},
\end{equation}
\noindent where $\sigma$ is a small positive constant set to $1 \times 10^{-5}$ for numerical stability. Under this path, $\lambda=0$ corresponds to the initial Gaussian noise latent, while $\lambda=1$ approaches the target latent representation. The corresponding target velocity is defined as
\begin{equation}
\boldsymbol{v}
=
\frac{d\boldsymbol{z}_{\lambda}}{d\lambda}
=
\boldsymbol{z}_1 - (1-\sigma)\boldsymbol{z}_{0}.
\end{equation}

The target latent $\boldsymbol{z}_1$ and the mixture latent $\boldsymbol{z}^{m}$ are obtained by encoding the target waveform $a$ and the mixture waveform $a^{m}$ using the VAE encoder:
\begin{equation}
\boldsymbol{z}_1 = f_{\text{VAE}}(a), \qquad
\boldsymbol{z}^{m} = f_{\text{VAE}}(a^{m}).
\end{equation}
In practice, the mixture latent $\boldsymbol{z}^{m}$ is concatenated with the intermediate latent $\boldsymbol{z}_{\lambda}$ along the channel dimension before being passed into the diffusion-based transformer~(DiT) generator, with the textual query embedding $\boldsymbol{E}$ used as language condition. The standard rectified flow objective is therefore written as
\begin{equation}
\mathcal{L}_{\text{RFM}}(\theta) =
\mathbb{E}_{\lambda, \boldsymbol{z}_1,\boldsymbol{z}_0}
\left\|
\mu\left([\boldsymbol{z}_{\lambda};\boldsymbol{z}^{m}], \boldsymbol{E}; \theta\right)
-
\boldsymbol{v}
\right\|^2
\end{equation}
\noindent where $[\cdot;\cdot]$ denotes channel-wise concatenation, $\mu(\cdot;\theta)$ denotes the DiT flow module, and $\boldsymbol{E}$ is the textual query embedding extracted by the text encoder.

\subsubsection{Self-Supervised Flow Matching~(Self-Flow)}
We apply Self-Flow to enhance semantic representation learning during flow training. The key idea is to construct two views of the same target latent with different levels of noise corruption. A student network receives a noisy view, in which different latent tokens are located at different flow times, while a teacher receives a uniform and cleaner view. The student is then trained to match an intermediate representation produced by the teacher.


As illustrated in the lower part of Figure~\ref{fig:overview}, the teacher is an EMA updated from the student. For the upper part of the Figure, the system applies the student as the final model for inference. Starting from a Gaussian latent $\boldsymbol{z}_0$, the model predicts the velocity field conditioned on $[\boldsymbol{z}_{\lambda};\boldsymbol{z}^{m}]$ and $\boldsymbol{E}$. The ordinary differential equation (ODE) solver~\cite{lu2022dpmsolver} is integrated from $\lambda=0$ to $\lambda=1$ to obtain the generated target latent $\hat{\boldsymbol{z}}_1$, which is then decoded by the VAE decoder to reconstruct the generated waveform.


During training, Self-Flow introduces an additional teacher-student representation learning objective on top of the standard RFM loss. To construct the student input, we first sample two flow times $\lambda_a$ and $\lambda_b$. Each latent token is independently assigned one of the two flow times by sampling a binary variable from Bernoulli distribution. Specifically, for the $i$-th token, its flow time is defined as
\begin{equation}
\lambda_i =
m_i \lambda_a + (1-m_i)\lambda_b,
\qquad
m_i\sim\operatorname{Bernoulli}(r_m),
\end{equation}
where $m_i\in\{0,1\}$ is a binary assignment variable sampled independently for each latent token. The variable $r_m$ controls the proportion of tokens assigned to each flow time and the student view is a collection of all tokens $ \boldsymbol{z}_{\boldsymbol{\lambda}}=\left\{z_{\lambda_i}^{(i)}\right\}_{i=1}^{N},$ where $N$ denotes the number of temporal latent tokens. The resulting student view therefore contains latent tokens corresponding to different flow times within the same sequence. In contrast, the teacher receives a uniform and cleaner view constructed using
\begin{equation}
\lambda_c = \max(\lambda_a,\lambda_b). 
\end{equation} 
Compared with the student input, $\boldsymbol{z}_{\lambda_c}$ contains a larger contribution from the target latent and a smaller contribution from Gaussian noise. Moreover, all teacher tokens share the same flow time, providing a more coherent representation target. As shown in the lower part of Figure~\ref{fig:overview}, the teacher network has the same architecture as the student network, but its parameters are updated as an exponential moving average of the student parameters. 

In both teacher and student branches, the mixture latent $\boldsymbol{z}^{m}$ is concatenated with the corresponding latent view before being sent to the DiT flow module, while the text embedding $\boldsymbol{E}$ is used as the semantic condition. This conditioning scheme allows the flow module to estimate the embedding of the target source ${\boldsymbol{z}}_1$ from the mixture-aware latent $\boldsymbol{z}^{m}$ and the textual query $\boldsymbol{E}$.

Following the cross-layer prediction design of Self-Flow, the deeper teacher layer provides a more contextualized target, while the earlier student layer is encouraged to acquire semantic representations at an earlier stage of the flow backbone. In this case, the loss used to align the Self-Flow representation is defined in terms of an intermediate student feature at the layer $l$ and an intermediate teacher feature at the layer $k$~($l<k$):
\begin{equation}
\resizebox{0.44\textwidth}{!}{$
\mathcal{L}_{\text{rep}}(\theta) =
\mathbb{E}
\left[
1 -
\cos\left(
h_{\theta}^{(l)}([{\boldsymbol{z}}_{\lambda};\boldsymbol{z}^{m}], \boldsymbol{E}),
\;
h_{\theta'}^{(k)}([\boldsymbol{z}_{\lambda_c};\boldsymbol{z}^{m}], \boldsymbol{E})
\right)
\right]
$}
\end{equation}
\noindent where $h_{\theta}^{(l)}(\cdot)$ and $h_{\theta'}^{(k)}(\cdot)$ denote the intermediate features projected from the DiT flow modules of the student and the teacher, respectively. The teacher parameters $\theta'$ are updated as the exponential moving average of the student parameters $\theta$ and are not directly optimized by back-propagation. The final training objective combines the original rectified flow matching loss and the Self-Flow representation learning loss:
\begin{equation}
\mathcal{L}_{\text{total}} = \mathcal{L}_{\text{RFM}} + \gamma \mathcal{L}_{\text{rep}},
\end{equation}
\noindent where $\gamma$ controls the weight of the representation objective. It is noted that the conventional uniformly noised view is retained for $\mathcal{L}_{\text{RFM}}$, while the dual-timestep student–teacher pair is used only for $\mathcal{L}_{\text{rep}}$.

\section{Datasets and Evaluation Benchmark}
\label{sec: dataset}

Following the data construction pipeline of FlowSep~\cite{flowsep}, we expand the training set to $5{,}650$ hours of audio clips. All waveforms are pre-processed and converted to mono and resampled to $16$ kHz before training and evaluation. In this section, we first present a detailed description of the datasets used to train FlowSep2, and then introduce the benchmarks used to evaluate separation performance.

\subsection{Training Datasets}

\subsubsection{VGGSound}
VGGSound~\cite{vggsound} is a large-scale audio dataset containing $200{,}000$ audio clips. Each clip has a duration of $10$ seconds and is annotated with one of 309 sound event labels. In this work, we use the original version of VGGSound, which contains $183{,}727$ audio-visual clips for training and $15{,}449$ clips for testing. 

\subsubsection{AudioCaps}
AudioCaps~\cite{audiocaps} is a publicly available audio captioning dataset consisting of 10-second audio clips paired with human annotated captions. AudioCaps contains $49{,}837$ training clips and $957$ testing clips. Each training clip is associated with a single caption, while each test clip has five captions.

\subsubsection{AudioSetCaps}
AudioSetCaps~\cite{bai2025audiosetcaps} is a large-scale audio-caption dataset built upon AudioSet, containing approximately $1.9$ million audio-caption pairs and representing one of the largest audio-language datasets. Unlike AudioSet, which only provides weak sound event labels, AudioSetCaps enriches the original recordings with synthetic natural language captions generated through an automated annotation pipeline. In this work, we collect $1.4$ million $10$-second audio clips from AudioSetCaps for training. 

\subsubsection{WavCaps}
WavCaps~\cite{wavcaps} is a large-scale audio captioning dataset. The captions are automatically generated with the assistance of large language models. The audio samples are collected from diverse sources, including FreeSound, BBC Sound Effects, SoundBible, and AudioSet. To maintain consistency with our 10-second training setup, we select audio samples shorter than $10$ seconds and use a subset of $400{,}000$ clips for training only. 

\begin{table}[t]
\centering
\caption{Summary of datasets used for training and evaluation in FlowSep2.}
\label{tab:dataset_setup}
\resizebox{0.48\textwidth}{!}{
\begin{tabular}{ccccc}
\toprule
Split & Dataset & Num. & Hours & Target Type  \\
\midrule
\multirow{4}{*}{\centering Train} 
& VGGSound & 183,727 & 510 & Label  \\
& AudioCaps & 49,837 & 140 & Caption  \\
& AudioSetCaps & 1,400,000 & 3,900 & Caption  \\
& WavCaps & 400,000 & 1,100 & Caption  \\
\midrule
\multirow{6}{*}{\centering Test}  
& VGGSound & 2,000 & 5.6 & Label  \\
& ESC-50  & 2,000 & 5.6 & Label  \\
& MUSDB  & 192 & 0.93 & Label \\
& AudioCaps  & 928 & 2.6 & Caption  \\
& DCASE-Synth & 3,000 & 8.3 & Caption  \\
& DCASE-Real & 100 & 0.3 & Caption  \\
\bottomrule
\end{tabular}
}
\end{table}

\subsection{Evaluation Benchmark}

We evaluate FlowSep2 on six different benchmarks. During evaluation, we ensure that the target and interference sources within each mixture do not share overlapping sound classes. It is also noted that all the samples of the AudioCaps test set are removed from the training data of AudioSetCaps and WavCaps. 

\subsubsection{VGGSound Benchmark}

For evaluation on VGGSound, we follow the benchmark construction protocol adopted in prior language-guided audio separation work~\cite{audiosep,flowsep}. Specifically, we first select $200$  audio clips as target sources. For every target clip, we randomly sample $10$ interference clips from the remaining test set to form target-interference pairs, resulting in a total of $2{,}000$ mixtures for evaluation. 

\subsubsection{AudioCaps Benchmark}

For AudioCaps evaluation, we use the first caption from the official test split, collecting $928$ audio-caption pairs. Each target audio clip is paired with another clip randomly sampled from the same test split to serve as the interference source. 


\subsubsection{ESC-50 Benchmark}

We evaluate the zero-shot separation performance for environmental sounds on ESC-50~\cite{esc-50}. ESC-50 consists of $2{,}000$ environmental audio recordings evenly distributed over $50$ semantic classes. For evaluation, we randomly select two audio clips from different categories, zero-pad each clip to $10$ seconds, and then mix them to form the mixtures for evaluation. 


\subsubsection{MUSDB18 Benchmark}

We evaluate the zero-shot performance of the proposed method on MUSDB18~\cite{musdb}, a widely used benchmark corpus for music separation. MUSDB18 contains $150$ full-track songs from different musical styles, including $100$ tracks for training and $50$ tracks for testing. Each song provides a stereo mixture together with four isolated source stems. For evaluation, we randomly extract a $10$-second segment from each track. For each segment, the original mixture is used as input, and each available source stem (e.g., drums, bass, and vocals) is treated as the target source, forming mixture–source pairs. To ensure data quality, we remove segments whose target stems are silent or contain negligible energy based on an amplitude threshold. After filtering, we obtain a total of $192$ valid waveform clips and form $142$ mixture–source pairs for evaluation. 

Unlike environmental sound benchmarks, MUSDB18 mainly contains music signals whose sources belong to closely related musical categories. Therefore, we only report CLAP$_A$ score and relevant subjective evaluation to assess whether the separated outputs correspond to the queried musical source and preserve perceptual audio quality.

\subsubsection{DCASE 2024 Task 9 Benchmark}

We evaluate the proposed model on the two official evaluation sets from DCASE 2024 Task 9~\footnote{https://dcase.community/challenge2024/task-language-queried-audio-source-separation}, namely DCASE-Synth (DE-S) and DCASE-Real (DE-R). These two benchmarks assess separation quality in both synthetically constructed mixtures and real-world overlapping acoustic scenes.

\paragraph{DCASE-Synth (DE-S)}
DE-S contains $3{,}000$ synthetic mixtures generated from $1{,}000$ source clips.  Each source clip is associated with a human-written text description, and the benchmark provides a corresponding textual query for identifying the target source within the mixture.

\paragraph{DCASE-Real (DE-R)}
DE-R consists of $100$ real-world recordings, each containing at least two overlapping sound sources captured in unconstrained acoustic environments. Following the official protocol, each DE-R recording is associated with two text queries.

\subsection{Evaluation Metrics}

Unlike discriminative separation systems that directly estimate time-frequency masks from the mixture signals, the proposed framework performs language-guided source separation through generative modeling in latent space. As a result, the separated outputs are not strictly aligned with the target reference signals in the temporal dimension. Traditional distortion-based source separation metrics, such as source-to-distortion ratio (SDR), are not well suited for evaluating the proposed system~\cite{flowsep}. To evaluate the separation performance, we adopt a set of objective and subjective metrics.  


\subsubsection{Fr\'echet Audio Distance (FAD)}

Fr\'echet Audio Distance (FAD) is a widely used metric for evaluating distribution-level similarity between generated audio and reference audio. A lower FAD score indicates that the separated audio distribution is closer to the target audio distribution, suggesting better perceptual realism and acoustic consistency.

\subsubsection{CLAP Score}

To evaluate semantic consistency between the separated audio and the language query, we use the CLAP score based on the CLAP model. Given the generated audio embedding $\mathbf{e}_a$ and the text embedding $\mathbf{e}_t$, the CLAP score is computed as the cosine similarity:

\begin{equation}
\text{CLAP}(\mathbf{e}_{a}, \mathbf{e}_{t}) =
\frac{\mathbf{e}_{a} \cdot \mathbf{e}_{t}}
{\max(\|\mathbf{e}_{a}\|\|\mathbf{e}_{t}\|, \epsilon)},
\end{equation}
where $\epsilon$ is a small constant introduced to avoid numerical instability, and a higher CLAP score indicates stronger semantic alignment between the generated audio and the text query. 

\subsubsection{CLAP$_A$ Score}

In addition to audio-text consistency, we further evaluate the similarity between the separated output and the target source itself using the CLAP$_A$ score. CLAP$_A$ measures the cosine similarity between the generated audio embedding $\mathbf{e}_{a}$ and the target audio embedding $\mathbf{e}_{\hat{a}}$:

\begin{equation}
\text{CLAP}_A(\mathbf{e}_{a}, \mathbf{e}_{\hat{a}}) =
\frac{\mathbf{e}_{a} \cdot \mathbf{e}_{\hat{a}}}
{\max(\|\mathbf{e}_{a}\|\|\mathbf{e}_{\hat{a}}\|, \epsilon)}.
\end{equation}

A higher CLAP$_A$ score indicates that the separated output is more similar to the target source in the learned embedding space, which is particularly useful to compare the similarity between the target and the generated waveform.

\subsubsection{AudioBox Aesthetics}

AudioBox Aesthetics~\cite{Audiobox_Aesthetics} is a recently proposed reference-free audio quality assessment framework that has demonstrated strong correlation with human perceptual ratings. It employs a pre-trained model to evaluate the quality of audio samples without using reference signals. In this work, we calculate the average score of two dimensions from AudioBox Aesthetics: Production Quality (PQ) and Content Usefulness (CU). PQ reflects the overall quality of the audio, including cleanness, fidelity, and the absence of artifacts, while CU measures the usefulness and meaningfulness of the audio content.

\begin{table}[t]
\centering
\caption{Experimental setups and model sizes, SF denotes Self-Flow and RFM is short for Rectified Flow-matching}
\label{tab:model-setups}

\resizebox{0.48\textwidth}{!}{%
\begin{tabular}{ccccc}
\toprule
Model & Training-Objectives & Encoder & Backbone & Params. (M) \\
\midrule

AudioSep       & Masking-Based   & STFT      & UNet & $238.6$  \\
SAM-Audio      & RFM   & VAE & DiT  & $2,934$ \\
FlowSep        & RFM  & VAE     & UNet & $265.5$ \\
\midrule
FlowSep2-UNet  & RFM  & \multirow{2}{*}{VAE}
                            & UNet & $352.1$ \\
FlowSep2-DDPM  & DDPM &      & DiT  & $559.5$ \\
\midrule
FlowSep2-RFM   & \multirow{3}{*}{RFM}
                       & VAE & \multirow{3}{*}{DiT} & $558.2$ \\
FlowSep2-RAE   &       & RAE &                         & $685.8$ \\
FlowSep2-REPA  &       & VAE &                         & $605.3$\\
\midrule
FlowSep2       & SF & VAE & DiT & $598.9$ \\
\bottomrule
\end{tabular}%
}
\end{table}

\begin{table*}[htbp]
\caption{Objective evaluation on LASS, where AC, VGG, ESC, MUC are short for AudioCaps~\cite{audiocaps}, VGGSound~\cite{vggsound}, ESC-50~\cite{esc-50}, and MUSDB18~\cite{musdb} respectively. FAD cannot be applied to unprocessed data as it calculates the difference between two groups of audio, while the target and mixed audio share the same audio events.} 
\centering
\small
\resizebox{0.95\textwidth}{!}{%
\begin{tabular}{ccccc|ccccc|ccccc}
\toprule
                     \multirow{3}{*}{Model}  & \multicolumn{4}{c}{FAD $\downarrow$} &  \multicolumn{5}{c}{CLAP Score $\uparrow$}& \multicolumn{5}{c}{CLAP$_{A}$ Score $\uparrow$}\\
 \cmidrule(lr){2-15}
&  AC & DE-S  & VGG & ESC  &  AC & DE-S  & DE-R& VGG & ESC  &  AC & DE-S  &  VGG & ESC & MUC \\
\midrule
Unprocessed
&    $-$    &  $-$  &  $-$   & $-$ & $31.7$ &$42.7$ &  $41.7$&    $33.6$    &  $38.6$   &  $54.4$   & $66.3$  &   $62.7$ &  $66.8$ & $50.2$    \\
LASS-Net~\cite{liu2022separate}
&    $5.09$    &  $1.83$  &  $3.09$   & $3.28$ & $33.9$ &$44.4$ &  $44.8$ &    $36.4$    &  $40.5$   &  $65.2$   & $72.1$ &   $64.5$ &  $75.6$ & $-$  \\
AudioSep~\cite{audiosep}
 &    $4.38$    &  $1.21$ & $2.30$     & $1.93$   & $33.6$ &$45.1$ &  $49.7$ &    $38.5$    &  $40.2$   &  $65.1$   & $74.4$ &  $67.4$ &  $76.0$  & $50.4$  \\
FlowSep~\cite{flowsep}
&  $2.86$   &  $0.90$  &   $2.06$     & $1.49$   & $41.9$ &$46.4$&  $50.3$ &    $39.5$    &  $42.2$   &  $76.7$   & $75.6$  &  $69.2$ &  $75.7$ & $54.1$\\
SAM-Audio~\cite{sam-audio}
&  $1.58$   &  $0.98$  &   $1.86$     & $1.20$   & $37.8$ &$43.4$&  $45.8$ &    $39.1$    &  $40.9$   &  $64.1$   & $71.3$  &  $70.5$ &  $78.6$ & $\textbf{62.8}$\\
\midrule
FlowSep2-M
& $\textbf{0.88}$ & $\textbf{0.72}$ & $\textbf{1.32}$ & $\textbf{1.15}$ 
& $\textbf{44.9}$ & $\textbf{47.5}$ & $\textbf{51.0}$ & $\textbf{42.0}$ 
& $\textbf{44.5}$ & $\textbf{80.5}$ & $\textbf{78.5}$ & $\textbf{72.0}$ 
& $\textbf{80.6}$ & $58.5$ \\
\bottomrule
\end{tabular}
}
\label{tab:objective_results}
\vspace{-0.4cm} 
\end{table*}

\subsubsection{SAJ Score}

To provide a more perceptually grounded evaluation for open-domain sound separation, we employ the SAM Audio Judge (SAJ), a reference-free automatic assessment model from SAM-Audio~\cite{sam-audio}. The SAJ model adopts a Transformer-based architecture and the system is trained to predict scores on a five-point Likert scale. Prior studies have shown that SAJ achieves strong correlation with human judgments across speech, music, and general sound domains, making it particularly suitable for evaluating language-guided universal sound separation.

Specifically, we apply the SAJ score by taking the average of four perceptual dimensions: Recall, Precision, Faithfulness, and Overall Quality. Recall measures whether all target sound events described by the query are successfully preserved in the separated output. Precision reflects the extent to which interfering non-target sounds are removed. Faithfulness evaluates whether the extracted target sounds remain acoustically similar to their corresponding instances in the original mixture. Overall Quality provides a holistic perceptual judgment of the final separated result.

\subsubsection{Subjective Evaluation}

We further conduct human subjective evaluations to assess the perceptual quality and semantic correctness of the separated audio. Following the evaluation protocol used in prior audio generation work~\cite{audioldm2}, we consider two complementary criteria: Overall Impression (OVL) and Audio-Text Relation (REL). For OVL, the question is: \textit{How would you rate the overall quality of this audio sample?} Responses were collected using a five-point Likert scale ranging from $1$ (bad) to $5$ (excellent). This metric reflects the overall perceptual quality of the separated audio, including fidelity, cleanness, and the absence of obvious artifacts. For REL, the question is: \textit{Does the generated audio correctly represent the same sound events compared to the target ground-truth?} This criterion evaluates whether the separated output is semantically consistent with the language condition and whether the intended target events are successfully extracted.

The subjective evaluation was conducted with ten human raters recruited through an open advertisement within the University of Surrey. None of the participants were involved in model development or the writing of this paper, and none of the authors participated in the evaluation. To encourage diverse perspectives and reduce potential bias, the participant pool included six raters with backgrounds in audio-related research and four raters from unrelated fields such as computer vision and robotics, thus providing both expert and non-expert perceptual assessments.

\section{Experimental Setting} 
\label{sec: experiment}

\subsection{Model Architecture Details}
For the text encoder, we adopt the encoder module from the pre-trained FLAN-T5 model~\cite{chung2022scaling-flan-t5}, which produces a sequence of text embeddings with a fixed length of 50 and a feature dimension of $1024$. For audio representation, instead of using the mel-spectrogram-based VAE employed in the original FlowSep system, we adopt the pretrained VAE from Stable-Audio~\cite{stableaudio}. This VAE directly operates on waveform signals and provides a compression ratio of approximately $320$. For a $10$-second audio clip, the encoder produces a latent representation with a temporal dimension of $512$ and a frequency dimension of $128$, which serves as the input to the generative backbone.
We evaluate three model scales of the proposed DiT backbone, denoted as FlowSep2-S, FlowSep2-M, and FlowSep2-L, with transformer depths $n_{\text{depth}} = 12$, $16$, and $24$, respectively. To further analyze the contribution of individual components, we evaluate several system variants:

\noindent\textbf{Generative backbone:}
FlowSep2-UNet replaces the DiT backbone with a UNet architecture following our previous FlowSep design~\cite{flowsep}.

\noindent\textbf{Training strategy:}
We compare different generative training formulations by adopting the traditional DDPM and the previous RFM, denoted as FlowSep2-DDPM and FlowSep2-RFM, respectively.

\noindent\textbf{Semantic Representation Alignment:}
Despite the proposed Self-Flow strategy, we further investigate two alternative strategies for semantic representation alignment. 

First, FlowSep2-RAE follows the idea of representation autoencoders \cite{rae} by replacing the original VAE latent space with semantic representations extracted by a frozen pretrained AudioMAE\cite{audiomae} encoder. Compared with conventional VAE latents, AudioMAE is expected to provide richer semantic representations while preserving acoustic information. To reconstruct audio from the semantic latent space, we directly adopt the pretrained diffusion-based decoder from SemantiCodec~\cite{liu2024semanticodec} without any additional fine-tuning.

Second, we introduce FlowSep2-REPA, which incorporates a semantic representation alignment objective inspired by REPA~\cite{repA}. Specifically, the hidden representation from the final DiT layer is first projected into the CLAP semantic embedding space and then aligned with the corresponding frozen CLAP text embedding using a cosine similarity loss. Unlike the proposed FlowSep2 with Self-Flow, this objective explicitly regularizes the internal representations of the flow model to preserve semantic information during training. Following the original REPA setting~\cite{repA}, the semantic alignment loss is enabled only after 500,000 training steps to avoid interfering with the early optimization of the flow matching objective.

These two variants investigate two complementary directions for improving semantic representation learning: replacing the latent representation itself (FlowSep2-RAE) and explicitly regularizing the latent representation during training (FlowSep2-REPA).

\subsection{Training and Inference Setup}
We collect $2,033,564$ audio clips for training and $8,220$ clips for evaluation. 
All models are trained on a single NVIDIA A100-80GB GPU for 4 million training steps. We use a batch size of 8 and optimize the model using the AdamW optimizer~\cite{loshchilov2017decoupled} with a learning rate of $5 \times 10^{-5}$. 
A linear warm-up schedule is applied over the first 10,000 steps.

\begin{table*}[tbp]
\caption{Subjective and human-aligned evaluation results on LASS, where AC, DE-S and DE-R denote AudioCaps~\cite{audiocaps}, DCASE-Synth and DCASE-Real respectively. SAJ (SAM Audio Judge), AA (AudioBox Aesthetics), REL and OVL are evaluated using a $1$–$5$ Likert scale. $\uparrow$ indicates that higher scores are better.}
\centering
\small
\resizebox{\textwidth}{!}{%
\begin{tabular}{cccc|cccc|cccc|cccc}
\toprule
\multirow{3}{*}{Model} 
& \multicolumn{3}{c}{SAJ $\uparrow$}
& \multicolumn{4}{c}{AA $\uparrow$}
& \multicolumn{4}{c}{REL $\uparrow$} 
& \multicolumn{4}{c}{OVL $\uparrow$} \\
\cmidrule(lr){2-16}
& AC & MUC & DE-S 
& AC & MUC & DE-S & DE-R
& AC & MUC & DE-S & DE-R 
& AC & MUC & DE-S & DE-R \\
\midrule

LASS-Net~\cite{liu2022separate}
& $2.57$ & $2.03$ & $2.68$ 
& $4.86$ & $5.27$ & $5.41$ & $4.89$
& $3.12$ & $2.87$ & $2.96$ & $3.59$
& $2.16$ & $2.45$ & $2.84$ & $3.88$ \\

AudioSep~\cite{audiosep}
& $2.84$ & $2.97$ & $2.93$
& $5.18$ & $5.94$ & $6.03$ & $5.57$
& $3.66$ & $3.99$ & $3.24$ & $3.93$
& $2.69$ & $3.85$ & $3.53$ & $4.02$ \\

SAM-Audio~\cite{sam-audio}
& $2.58$ & $\textbf{3.76}$ & $2.91$
& $\textbf{5.56}$ & $6.48$ & $\textbf{6.14}$ & $5.79$
& $3.54$ & $\textbf{4.35}$ & $3.24$ & $3.85$
& $2.69$ & $\textbf{4.44}$ & $3.75$ & $4.26$ \\

FlowSep~\cite{flowsep}
& $2.73$ & $3.08$ & $3.02$
& $5.36$ & $6.07$ & $5.83$ & $5.28$
& $4.08$ & $3.71$ & $\textbf{3.62}$ & $4.11$
& $3.98$ & $3.84$ & $3.72$ & $4.26$ \\
\midrule

FlowSep2-M
& $\textbf{2.96}$ & $3.39$ & $\textbf{3.47}$
& $5.49$ & $\textbf{6.67}$ & $6.08$ & $\textbf{5.86}$
& $\textbf{4.13}$ & $3.91$ & $\textbf{3.62}$ & $\textbf{4.16}$
& $\textbf{4.09}$ & $4.02$ & $\textbf{3.88}$ & $\textbf{4.28}$ \\

\bottomrule
\end{tabular}
}
\label{tab:subjective_results}
\end{table*}

\section{Results and Analysis}
\label{sec: results}
\begin{figure*}[t]
    \centering
    \includegraphics[width=0.95\textwidth]{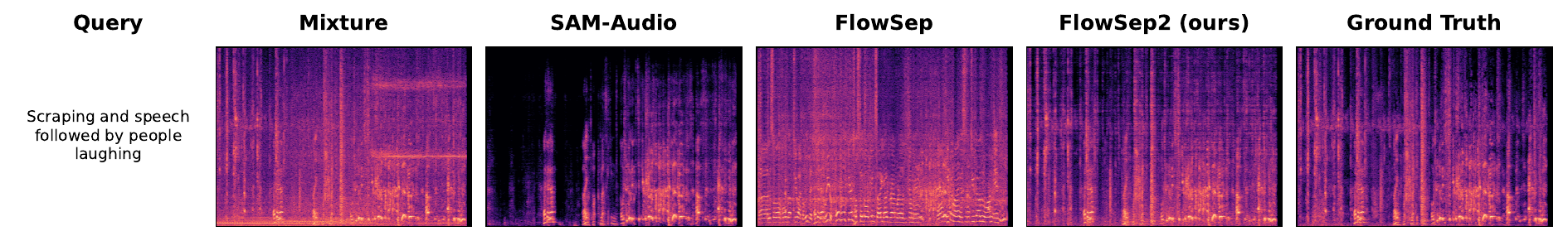}
    \caption{A case study comparing separation results from different methods with the ground truth on the AudioCaps test set.}
    \label{fig:qualitative_comparison}
\end{figure*}

This section presents the key results in both objective and subjective metrics, as shown in Tables~\ref{tab:objective_results} and~\ref{tab:subjective_results}, respectively. We can observe that FlowSep2 consistently achieves state-of-the-art performance in most cases among both discriminative and generative baselines, demonstrating enhanced separation performance and perceptual audio quality. In addition, Figure~\ref{fig:qualitative_comparison} provides a comparison of the separation results from different methods. In this example, FlowSep2 reconstructs the target spectrogram more faithfully while retaining less visible interference from non-target sources.

\subsection{Results on Objective Metrics}
Compared with FlowSep, the proposed FlowSep2 obtains the lowest FAD scores across all the evaluated datasets, including AudioCaps, DCASE-Synthetic, VGGSound, and ESC-50. In particular, FlowSep2 reduces the FAD score from $2.86$ to $0.88$ on AudioCaps and from $2.06$ to $1.32$ on VGGSound, indicating that the distribution of the separated audio becomes significantly closer to that of the target audio. FlowSep2 also achieves the highest CLAP Score on all evaluation datasets, including synthetic and real-world DCASE benchmarks. Specifically, FlowSep2 reaches $44.9$ on AudioCaps and $51.0$ on DCASE-Real, outperforming other strong baselines. These results indicate a stronger semantic alignment between the separated audio and the language query, suggesting that the proposed semantic-enhanced latent representations and improved generative modeling effectively improve language-guided separation capability. In terms of the CLAP$_A$ Score, which measures the similarity between separated audio and target audio, FlowSep2 also achieves the highest overall performance on most benchmarks. Compared to FlowSep, its CLAP$_A$ Score has improved from $76.7$ to $80.5$ on AudioCaps and from $69.2$ to $78.5$ on VGGSound. 

On MUSDB, SAM-Audio achieves the highest score, which is expected because it is trained on large-scale music datasets and is therefore more directly optimized for music source separation. This makes SAM-Audio a strong state-of-the-art system for music separation. In contrast, FlowSep2 is trained exclusively on general sound datasets due to the limited availability of suitably licensed large-scale music datasets, and the system is evaluated on MUSDB entirely in a zero-shot setting. Despite this, FlowSep2 still achieves the best overall performance on general LASS benchmarks. These results suggest that FlowSep2 provides stronger general-purpose language-guided separation capability and better cross-domain generalization beyond music-specific separation.

\subsection{Results on Subjective and Human-aligned Metrics}

We further evaluate the perceptual quality of the separated audio, as shown in Table~\ref{tab:subjective_results}. FlowSep2 achieves the best overall performance across most evaluation settings, indicating that the proposed model not only improves objective metrics but also produces perceptually better and more semantically aligned results.

On AudioCaps and DCASE-Synthetic, FlowSep2 obtains the highest SAJ scores, reaching $2.96$ and $3.47$, respectively. This suggests that the separated sources are more consistent with the given language queries. FlowSep2 also improves over FlowSep on all SAJ benchmarks, showing that the improved semantic representation and backbone strengthen separation tasks. For AudioBox Aesthetics, which mainly reflects the overall perceptual quality of generated audio, FlowSep2 achieves SoTA results across most benchmarks. It obtains the best scores on AudioCaps, MUSDB, and DCASE-Real, indicating that FlowSep2 can generate high-quality separated clips with better perceptual naturalness and audio fidelity.

In terms of REL and OVL, FlowSep2 also achieves the highest scores on most benchmarks. Compared with FlowSep, FlowSep2 improves OVL from $3.98$ to $4.09$ on AudioCaps and from $3.72$ to $3.88$ on DCASE-Synthetic, demonstrating better perceptual quality and target-source relevance. The result on MUSDB is consistent with the objective evaluation, where SAM-Audio shows stronger music separation performance, while FlowSep2 remains competitive in the zero-shot setting and achieves stronger overall results. The subjective evaluation illustrates that FlowSep2 improves both perceptual audio quality and semantic faithfulness, while maintaining strong generalization across environmental sound, music, and real-world separation scenarios.

\section{Ablation Studies}
\label{sec: ablation}
This section uses ablation studies to demonstrate the choices made in semantic representation, training objectives, backbone architecture and model scales.

\begin{table*}[t]
\caption{
Ablation study on different design choices in FlowSep2, comparing backbone architectures (Backbone), diffusion training objectives (Dif.), semantic representation alignment strategies (Sem.), and model depths. Within each ablation group, only the modified design choices are highlighted in black, while unchanged settings are shown in gray. Performance values that improve upon the corresponding baseline (the first row of each ablation group) are underlined, and the best results are highlighted in bold.
}
\label{tab:ablation_results}
\centering
\small

\resizebox{0.99\textwidth}{!}{%
\begin{tabular}{cccc|cccc|ccccc|ccccc}
\toprule
\multirow{2}{*}{Backbone}
& \multirow{2}{*}{Dif.}
& \multirow{2}{*}{Sem.}
& \multirow{2}{*}{Depth}
& \multicolumn{4}{c|}{FAD $\downarrow$}
& \multicolumn{5}{c|}{CLAP Score $\uparrow$}
& \multicolumn{5}{c}{CLAP$_A$ Score $\uparrow$} \\
\cmidrule(lr){5-8}
\cmidrule(lr){9-13}
\cmidrule(lr){14-18}
&&&
& AC & DE-S & VGG & ESC
& AC & DE-S & DE-R & VGG & ESC
& AC & DE-S & VGG & ESC & MUC \\

\midrule

\textbf{DiT}
& \textbf{RFM}
& \textbf{Self-Flow}
& \textbf{16}
& $0.88$ & $0.72$ & $1.32$ & $1.15$
& $44.9$ & $47.5$ & $51.0$ & $42.0$ & $44.5$
& $80.5$ & $78.5$ & $72.0$ & $80.6$ & $58.5$ \\
\midrule

\textbf{UNet}
& \graytext{RFM}
& \graytext{-}
& \textbf{18}
& $1.68$ & $1.03$ & $1.85$ & $1.58$
& $40.5$ & $45.4$ & $49.4$ & $39.1$ & $42.1$
& $74.5$ & $75.8$ & $70.1$ & $77.1$ & $54.8$ \\

\graytext{DiT}
& \textbf{DDPM}
& \graytext{-}
& \graytext{16}
& $1.92$ & $1.02$ & $1.99$ & $1.61$
& $42.8$ & $45.8$ & $49.9$ & $39.8$ & $42.0$
& $72.6$ & $75.0$ & $68.4$ & $77.5$ & $53.5$ \\

\graytext{DiT}
& \textbf{RFM}
& \graytext{-}
& \graytext{16}
& $1.38$ & $0.81$ & $1.81$ & $1.47$
& $42.3$ & $46.9$ & $50.5$ & $40.8$ & $43.0$
& $78.6$ & $76.3$ & $70.9$ & $79.1$ & $56.9$ \\

\midrule

\graytext{DiT}
& \graytext{RFM}
& \textbf{RAE}
& \graytext{16}
& $\underline{0.86}$ & $0.79$ & $\underline{1.32}$ & $1.19$
& $44.5$ & $\underline{47.6}$ & $50.8$ & $\underline{42.1}$ & $\underline{44.6}$
& $\underline{80.8}$ & $78.0$ & $\underline{72.5}$ & $\underline{\textbf{80.6}}$ & $57.3$ \\

\graytext{DiT}
& \graytext{RFM}
& \textbf{REPA}
& \graytext{16}
& $0.99$ & $0.80$ & $1.48$ & $1.25$
& $44.1$ & $47.5$ & $50.5$ & $42.0$ & $44.4$
& $80.2$ & $78.2$ & $71.8$ & $80.4$ & $58.3$ \\

\midrule

\graytext{DiT}
& \graytext{RFM}
& \graytext{Self-Flow}
& \textbf{12}
& $1.09$ & $0.95$ & $1.35$ & $1.17$
& $43.5$ & $47.2$ & $51.0$ & $42.0$ & $44.1$
& $77.6$ & $77.6$ & $72.0$ & $80.4$ & $58.4$ \\

\graytext{DiT}
& \graytext{RFM}
& \graytext{Self-Flow}
& \textbf{24}
& $\underline{\textbf{0.84}}$ & $\underline{\textbf{0.71}}$
& $\underline{\textbf{1.28}}$ & $\underline{\textbf{1.14}}$
& $\underline{\textbf{45.2}}$ & $\underline{\textbf{48.9}}$
& $\underline{\textbf{51.2}}$ & $\underline{\textbf{42.3}}$
& $\underline{\textbf{44.5}}$
& $\underline{\textbf{81.3}}$ & $\underline{\textbf{78.9}}$
& $\underline{\textbf{72.4}}$ & $80.5$ & $\underline{\textbf{59.8}}$ \\

\bottomrule
\end{tabular}%
}
\end{table*}

\subsection{Semantic Representation Alignment}

As shown in Table~\ref{tab:ablation_results}, we first investigate different representation learning strategies, including VAE, RAE, and REPA. RAE-based representations lead to the best performance over the RFM baseline, reducing FAD from $1.38$ to $0.86$ on AudioCaps and from $1.81$ to $1.32$ on VGGSound. REPA also improves over the RFM baseline, showing that explicit representation alignment can enhance semantic learning during flow matching.

\begin{figure}[htbp]
    \centering
    \includegraphics[width=1.0\linewidth]{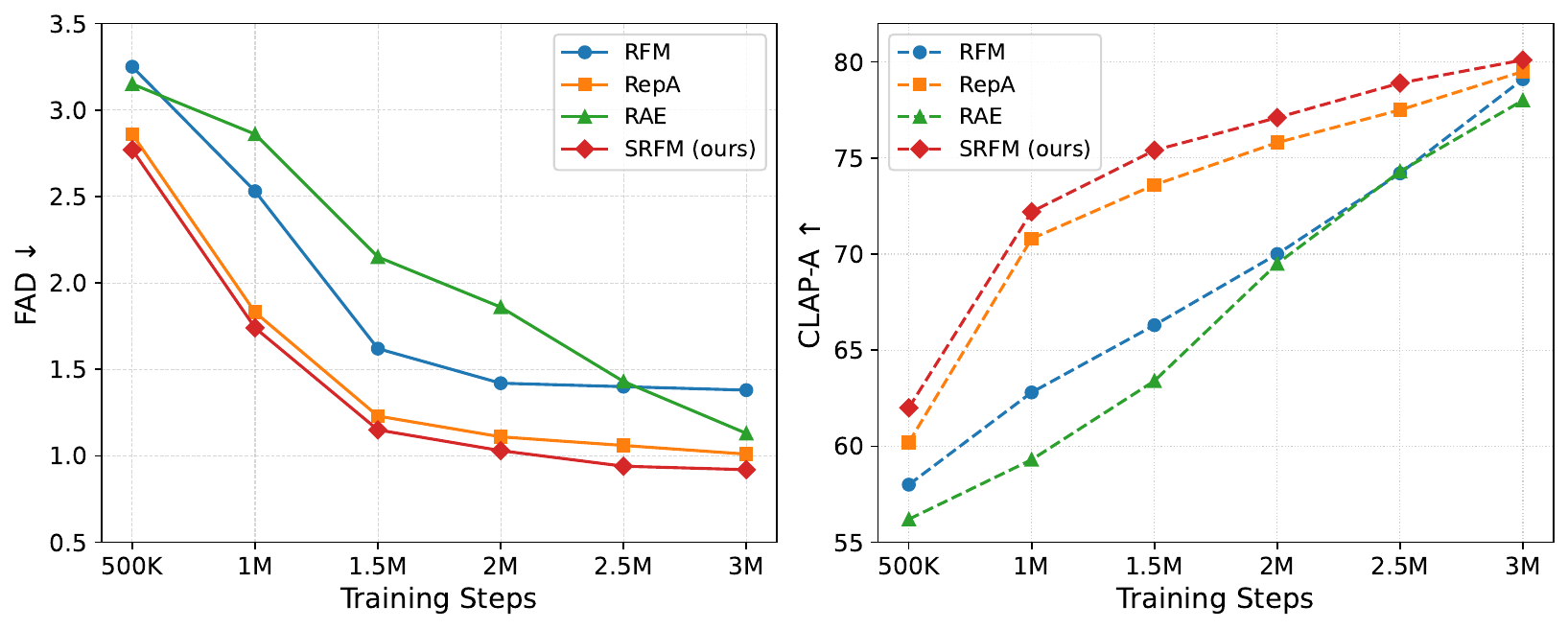}
    \caption{The comparison between different semantic-enhancement structures on different training steps. SRFM is short for Self-Flow }
    \label{fig:semtic_structure}
\end{figure}

In addition, the training curves in Figure~\ref{fig:semtic_structure} further show that semantic alignment strategies improve both convergence speed and overall performance. Although the standard RFM baseline converges reasonably well in about $2$M training steps, its performance remains below the semantic-enhanced models, suggesting that semantic structure learning is important for high-quality language-guided separation. REPA and Self-Flow~(SRFM) converge faster, reaching strong performance in about $1$M to $1.5$M training steps. In contrast, RAE converges more slowly, which may be attributed to the use of a larger and complex AudioMAE-based representation encoder, leading to a more challenging optimization process. While RAE achieves stronger final performance on several metrics, Self-Flow provides a better trade-off between convergence efficiency and overall separation quality.

Overall, these results demonstrate that semantic representation is critical for improving both semantic controllability and generation quality in source separation. Compared with existing semantic enhancement strategies, the proposed Self-Flow achieves competitive performance while maintaining substantially better training efficiency, making it a more practical and scalable solution for large-scale language-guided audio separation.

\subsection{Training Objectives}

We compare different training objectives between DDPM and rectified flow matching. As shown in Table~\ref{tab:ablation_results}, replacing conventional DDPM training with rectified flow matching consistently improves both generation quality and semantic alignment across all evaluation benchmarks.

Compared with FlowSep2-DDPM, FlowSep2-RFM achieves significantly lower FAD scores on all datasets. These results indicate that rectified flow matching produces cleaner and more realistic separated audio distributions than diffusion-based training. Semantic consistency is also substantially improved when using RFM training. FlowSep2-RFM consistently achieves higher CLAP score and CLAP$_A$ score across nearly all datasets. These results suggest that RFM improves the alignment between generated audio and language conditions while preserving the target source characteristics more effectively.

\subsection{Backbone Architecture}

We investigate the effect of different backbone architectures and model scales in FlowSep2. As shown in Table~\ref{tab:ablation_results}, replacing the conventional UNet backbone with a DiT-based architecture consistently improves both audio quality and semantic alignment across all evaluation benchmarks.

Compared with FlowSep2-UNet, the DiT-based FlowSep2-RFM achieves substantially lower FAD scores and higher CLAP-related metrics on most datasets. These results suggest that transformer-based latent generation is more effective than convolutional UNet architectures for modeling long-range semantic dependencies in language-guided source separation.

\subsection{Model Scaling}
We study the scalability of DiT backbones using three model sizes: FlowSep2-S~($12$ layers), FlowSep2-M~($16$ layers), and FlowSep2-L~($24$ layers). The results show consistent performance improvements as model capacity increases. FlowSep2-M already achieves strong performance across all benchmarks, outperforming most baseline systems in both FAD and semantic similarity metrics. FlowSep2-L provides additional improvements, achieving the best overall results on most objective metrics, including the lowest FAD and highest CLAP Score values in multiple datasets.

\section{Conclusion and Future Works}
\label{sec: conclusion}
In this paper, we presented FlowSep2, a scalable extension of our previous FlowSep framework with improved semantic representation alignment. In particular, the proposed system introduces semantic representation alignment into flow matching, improving both training convergence and separation performance. We further adopt a DiT backbone with an end-to-end waveform encoder and investigate model scaling, showing that larger transformer-based generators provide stronger global modeling capacity for complex acoustic scenes. Trained on more than $5{,}000$ hours of audio-language data, FlowSep2 also achieves improved open-domain generalization across diverse separation tasks, providing a generative alternative to conventional discriminative masking-based approaches.

Experimental results on multiple benchmarks demonstrate that FlowSep2 consistently outperforms previous generative and discriminative baselines on both objective and subjective evaluations. The ablation studies further validate the effectiveness of rectified flow matching, semantic representation learning, and DiT model scaling. For future work, we plan to further improve FlowSep2 by incorporating more music-related training data, such as stem-level music separation datasets, to enhance its performance on music source separation while maintaining its general-purpose separation capability. We will also explore richer conditioning modalities for separation, including spatial cues and visual information, which can provide additional source location and event-level context for more precise source separation in complex acoustic scenes. Overall, this work demonstrates the potential of generative frameworks for language-guided sound separation, providing a scalable foundation for future general-purpose CASA systems.

\section*{Acknowledgments}
\noindent
This research was partly supported by a research scholarship from the China Scholarship Council~(CSC), funded by British Broadcasting Corporation Research and Development~(BBC R\&D), Engineering and Physical Sciences Research Council~(EPSRC) Grant EP/T019751/1 ``AI for Sound'', and a PhD scholarship from the Centre for Vision, Speech and Signal Processing~(CVSSP), University of Surrey. 
For the purpose of open access, the authors have applied a Creative Commons Attribution~(CC BY) license to any Author Accepted Manuscript version arising. 

\bibliography{reference}
\bibliographystyle{IEEEtran}

\vfill

\end{document}